\documentclass[9pt,conference]{IEEEtran}
\usepackage{dcase2025}
\usepackage{gensymb}
\usepackage[defaultcolor=red]{changes}

\usepackage{bm} 

\usepackage{dcase2025,amsmath,graphicx,url,times,booktabs, tabularx}

\title{Region-Specific Audio Tagging for Spatial Sound}

\name{
\begin{tabular}{c}
Jinzheng Zhao$^{1}$,
      Yong Xu$^{2}$,
      Haohe Liu$^{1}$,
      Davide Berghi$^{1}$,
      Xinyuan Qian$^{3}$, \\
      Qiuqiang Kong$^{4}$,
      Junqi Zhao$^{1}$,
      Mark D. Plumbley$^{1}$,
      Wenwu Wang$^{1}$
      \end{tabular}
      \thanks{This research was supported by Tencent AI Lab Rhino-Bird Gift Fund and University of Surrey. This work was also supported by the Engineering and Physical Sciences Research Council [grant numbers EP/T019751/1, EP/Y028805/1]. For the purpose of open access, the authors have applied a creative commons attribution (CC BY) licence to any author accepted manuscript version arising. The codes and dataset are available at https://github.com/KawhiZhao/AudioTagging.}
      }
\address{$^{1}$Centre for Vision, Speech and Signal Processing (CVSSP), University of Surrey, UK  \\
$^{2}$Tencent AI Lab, Bellevue, WA, USA\\
$^{3}$Department of Computer Science and Technology, University of Science and Technology Beijing, China\\
$^{4}$The Chinese University of Hong Kong (CUHK)
}

\begin{document}

\maketitle

\begin{abstract}
    
Audio tagging aims to label sound events appearing in an audio recording. In this paper, we propose region-specific audio tagging, a new task which labels sound events in a given region for spatial audio recorded by a microphone array. The region can be specified as an angular space or a distance from the microphone. We first study the performance of different combinations of spectral, spatial, and position features. Then we extend state-of-the-art audio tagging systems such as pre-trained audio neural networks (PANNs) and audio spectrogram transformer (AST) to the proposed region-specific audio tagging task. Experimental results on both the simulated and the real datasets show the feasibility of the proposed task and the effectiveness of the proposed method. Further experiments show that incorporating the directional features is beneficial for omnidirectional tagging.
\end{abstract}

\section{Introduction}

The task of audio tagging aims to identify the sound events present in a sound clip. This topic has been a popular area in audio processing, playing an important role in applications, such as audio classification \cite{stowell2015detection} and information retrieval \cite{wold1996content}. In the current task setting, in general, all sound events are labeled regardless of the location of the sound events. In surveillance applications, however, sound events located in a specific spatial region may be more important than others. Therefore, it is of practical interest to study the problem of region-specific audio tagging, i.e., labelling audio events that are present in a specific spatial region. Tagging sound events according to the location can help the separation and make people focus on sound from a specific region. In addition, region-specific audio tagging allows people to attach importance to sounds, e.g. a warning from behind, thereby improving the safety.

In the traditional audio tagging task, deep learning based methods are popular choices. For a convolutional neural network (CNN) based system, pretrained audio neural networks (PANNs) \cite{kong2020panns} is a model pretrained on AudioSet \cite{gemmeke2017audio}, and shows promising performance on audio pattern recognition tasks such as audio tagging and acoustic scene classification. The system of pretraining, sampling, labeling, and aggregation (PSLA) \cite{gong2021psla} employs EfficientNet as the backbone and improves the model performance by ImageNet pretraining, balanced sampling and model aggregation. For a transformer-based \cite{waswani2017attention} system, the audio spectrogram transformer (AST) \cite{gong2021ast} follows the vision transformer \cite{dosovitskiy2020image}, and takes mel-spectrogram patches as input. The AST model can outperform PANNs and PSLA, when built with a large amount of data. The above methods only use spectral features. In \cite{xu2017convolutional}, both spectral and spatial audio features are used with a gated CNN architecture. The incorporation of audio features further improves the model performance. 
In addition to the CNN and transformer-based methods, there are also methods based on graph neural network (GNN), such as the work proposed in \cite{singh2024atgnn}, which leverages several GNNs to model the relationships between different feature patches and different labels for tagging.
In our work, we extend these state-of-the-art models to the region-specific audio tagging problem. 

Our proposed task is inspired by region-specific speech processing \cite{chen2018multi, gu2020multi, gu2024rezero} including automatic speech recognition (ASR) and speech separation for a given direction or distance. 
Beamforming aims to locate the signal from a given direction by attenuating unwanted sound sources. Traditional methods like the minimum variance distortionless response (MVDR) \cite{griffiths1982alternative} beamformer can minimize the total signal power and maintain a distortionless response for a given direction.
Recently, deep learning based methods are also used for region-specific automatic speech recognition and separation. 
For region-specific ASR, in \cite{chen2018multi}, long short term memory (LSTM)-based architecture is used for multi-channel overlapped ASR with the input of the concatenation of spectral, spatial and angle features. Directional features are proposed as the angle features, encoding the information for regions of interest.
For region-specific speech separation, in \cite{gu2020multi}, directional features are used as a condition to indicate the speaker for multi-modal target speech separation. In \cite{gu2024rezero}, a multi-channel band-split recurrent neural network (RNN) model is proposed for angular-query, spherical-query and conical-query based speech separation. 
In addition, the field of view (FOV) feature is proposed in \cite{yu2023deep} for audio zooming. The FOV feature can represent the property of an angular space while directional features can represent the property of an azimuth. 

\begin{figure}[tbp]
  \centering
  \includegraphics[width=\columnwidth]{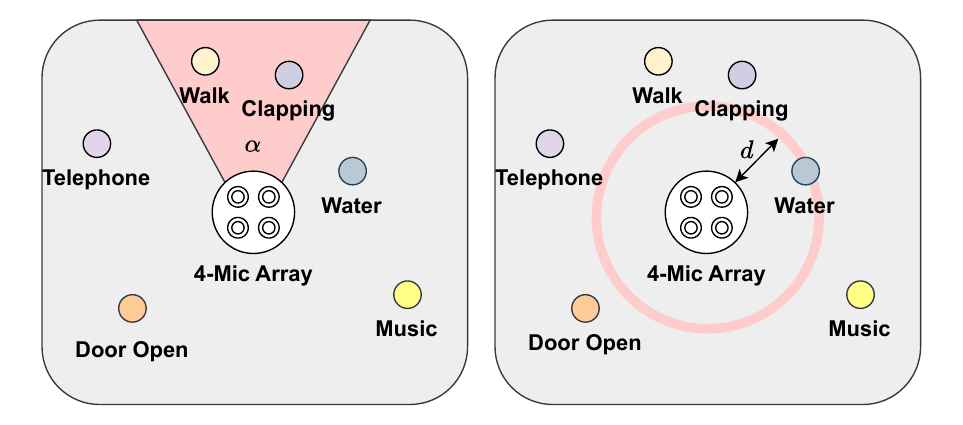}
  \caption{The task scenario of region-specific audio tagging. Left: Query by horizontal angular regions. Right: Query by distance.}
  \label{scene}
\end{figure}
\begin{figure*}[tbp]
  \centering
  \includegraphics[width=0.95\linewidth]{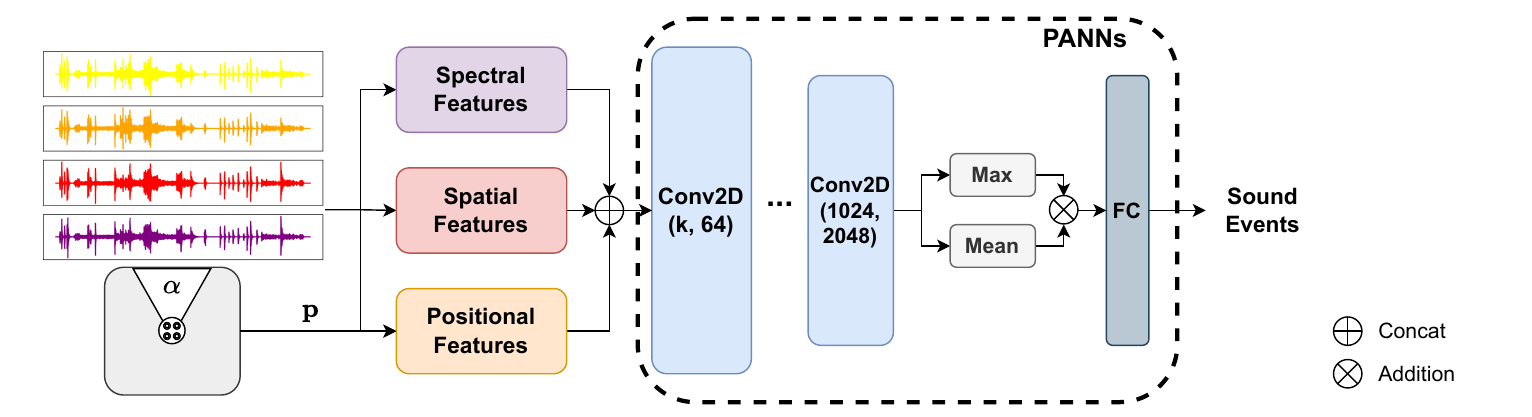}
  \caption{The proposed multi-channel PANNs for region-specific audio tagging.}
  \label{model}
\end{figure*}
\section{Proposed Method}
\label{method}
Inspired by the previous work in region-specific speech processing, we study a new task, i.e. region-specific audio tagging, and explore the use of directional features and FOV features for this task. Following previous work \cite{gu2024rezero}, we explore this task in two paradigms, query by horizontal angular regions and query by distance, as shown in Figure \ref{scene}. We examine three types of features, i.e. spectral, spatial, and positional features. The spectral features are used to identify sound sources. Spatial features can give information for all potential sound events and positional features are used to describe a potential region-specific sound event.

Our contributions can be summarized as follows. Firstly, we propose, for the first time, the region-specific audio tagging task and build a benchmark. Secondly, we extend the state-of-the-art audio tagging models for this task and study the performance achieved by a different combination of spectral, spatial and positional features. Finally, we extend the proposed method to omnidirectional audio tagging, which aims to tag the events in the whole space instead of a region. Experimental results show that the proposed fixed-region and location-aware system are superior to the omnidirectional tagging systems.

\subsection{Problem Definition}
Given a four-channel audio in tetrahedral microphone array (MIC) format $\mathbf{x} \in \mathbb{R}^{4 \times L}$ (where $L$ denotes the audio length) and a given area $\mathbf{p}$, in the form of angular range $[\theta_{\text{begin}}, ..., \theta_{\text{end}}]$ or distance $d$ from the microphone array to the sound events, the objective of region-specific audio tagging is to identify the sound sources located in $\mathbf{p}$. It is worth noting that the proposed new task can be achieved by a beamformer and an audio tagging model. However, this cascaded format is not end-to-end and the audio tagging model can only learns the semantic representation.

\subsection{Overall System Architecture}
To this end, we present a system, as shown in Figure \ref{model}. The system is composed of the modules including feature extraction and audio tagging. The features indicate the region of interests while the tagging model predicts the sound events. Compared to previous tagging systems, the proposed system predicts the sound events in the given region specified by a user, instead of all events present in microphone recordings. Compared to the cascaded model combining a beamformer and an audio tagging model, the proposed model can learn both semantic and spatial information simultaneously.

\subsection{Extraction of Features}
Following previous work for region-specific speech separation \cite{gu2020multi, gu2024rezero}, we also use the combination of spectral, spatial and positional features. 
\subsubsection{Spectral Features}
Spectral features can help classify the audio events. Here we use logarithm power spectrum (LPS) $\mathbf{L}$ of the first-channel audio $\mathbf{x}$. $\mathbf{L}$ is calculated based on short-time Fourier
transform (STFT) $\mathbf{X}$ of $\mathbf{x}$, as follows, 
\begin{equation}
    \mathbf{L} = \mathrm{log}(|\mathbf{X}|^{2})
\end{equation}
where $\mathbf{X} \in \mathbb{R}^{T \times F}$ and $\mathbf{L} \in \mathbb{R}^{T \times F}$, and $T$ and $F$ are the number of time frames and frequency bins, respectively. 

\subsubsection{Spatial Features}
Spatial features are useful for localizing sound events \cite{berghi2023audio}. Here, we explore two features, generalized cross-correlation phase transform (GCCPHAT) $\mathbf{G}$ and inter-channel phase difference (IPD) $\mathbf{I}$.

\textbf{GCCPHAT} is widely used in sound source localization \cite{zhao2023audio} and speaker tracking \cite{zhao2024attention}, which estimates the time difference of arrival between a pair of microphones, as follows,
\begin{equation}
    \mathbf{G}^{n} (\tau) = \int_{-\infty}^{+\infty} \frac{\mathbf{X}^{n_{1}} (f) \mathbf{X}^{n_{2} *} (f)}{|\mathbf{X}^{n_{1}} (f) \mathbf{X}^{n_{2} *} (f)|} e ^{i 2 \pi f \tau} d f
\end{equation}
where $n = (n_{1}, n_{2})$ indexes the microphone pair, $*$ denotes complex conjugate, $f$ denotes the frequency, and $\tau$ denotes the time delay. It is normalized to mitigate the impact of changing amplitudes. We stack GCCPHAT from selected pairs as the spatial features. 

\textbf{IPD} is calculated as the phase difference between two audio channels. 
\begin{equation}
    \mathbf{I}^{n} = \mathrm{angle}(\mathbf{X}^{n_{1}}) - \mathrm{angle}(\mathbf{X}^{n_{2}})
\end{equation}
where $\mathrm{angle}(\cdot)$ denotes the phase of a signal. IPD shows effectiveness in multi-channel speech separation \cite{gu2020multi}. IPDs from selected microphone pairs are stacked. 

\subsubsection{Positional Features}
In this section, we discuss the extraction of positional features from the angular region or a distance, respectively. 

\textbf{Angular Region} For the horizontal angular areas from $-180 \degree$ to $180 \degree$, we detect the sound events within the region $\Theta=[\theta_{\text{begin}}, ..., \theta_{\text{end}}]$ by attenuating events from the unselected areas. We choose the FOV feature proposed in \cite{yu2023deep} for this task.

To calculate the FOV feature, we first calculate directional features (DF) $\mathbf{D} \in \mathbb{R}^{T \times F}$ for each angle in $\Theta$ following \cite{wang2018spatial}, as follows,
\begin{equation}
   \mathbf{D}(\theta) = \sum_{n} \cos( \mathbf{I}^{n} - \mathbf{P}^{n}(\theta))
\end{equation}
\begin{equation}
    \mathbf{P}^{n}(\theta) = 2\pi f \phi^{n} \cos(\theta) f_{s} / c
\end{equation}
where $\mathbf{P} ^ {n}$ is the target-dependent phase difference calculated at the $n$-th microphone pair with frequency $f$, $\phi^{n}$ is the distance between the $n$-th microphone pair, $f_{s}$ is the sampling rate, and $c$ is the sound velocity.
Then, the FOV feature in the view of $\Theta$ is calculated as:
\begin{equation}
    \mathbf{F}_{\text{in}} = \mathrm{max}(\mathbf{D}(\theta)), \theta \in \Theta
\end{equation}
The feature out of the view $\Theta$ is calculated as: 
\begin{equation}
    \mathbf{F}_{\text{out}} = \mathrm{max}(\mathbf{D}(\theta)), \theta \notin \Theta
\end{equation}
Finally, the FOV feature $\mathbf{F} \in \mathbb{R}^{T * F}$ is defined as:
\begin{equation}
    \mathbf{F} = 
    \begin{cases}
\mathbf{F}_{\text{in}}, & \text{if }  \mathbf{F}_{\text{in}} > \mathbf{F}_{\text{out}}, \\
-1, & \text{if } \mathbf{F}_{\text{in}} \leq \mathbf{F}_{\text{out}}.
\end{cases}
\end{equation}

\begin{figure}[tbp]
  \centering
  \includegraphics[width=\columnwidth]{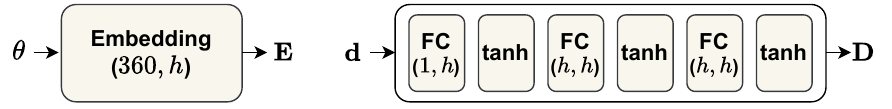}
  \caption{The learning-based method for obtaining angle features (left) and distance features (right), where $h$ is the hidden dimension and FC denotes the fully connected layer.}
  \label{distance}
\end{figure}

\textbf{Distance} For obtaining a distance feature given a distance $d$ from the microphone array to the sound events, we use learning-based methods following \cite{gu2024rezero}, as shown in the right part of Figure \ref{distance}. The distance feature is repeated to match the time dimension of spectral or spatial features. 

\subsection{Model}
We extended PANNs for region-specific multi-channel audio tagging, as shown in Figure \ref{model}. The spectral, spatial and positional features are concatenated on the channel dimension, with $k$ being the number of concatenated channels. PANNs takes the concatenated features and employs a sequence of convolutional layers to downsize the time-frequency dimension while increasing the channel dimension from $k$ to 2048. Then the time-frequency dimension is pooled and the channel dimension is retained. Finally, the system outputs the probabilities of the sound event classes. The binary cross-entropy loss is used to optimize the model.

We also extended the other two models, PSLA \cite{gong2021psla} and AST \cite{gong2021ast}, to the proposed task. For PSLA, the EfficientNet backbone is trained from scratch with 2560 hidden dimension in the multi-head attention module. For AST, we use the \textit{small224}\footnote{https://github.com/YuanGongND/ast/blob/master/src/models/ast\_models.py} version. 

\section{Experiment Setup}
\label{setup}
\subsection{Dataset}
We use the STARSS23 dataset \cite{shimada2024starss23} for the proposed task and this dataset is also used in DCASE 2024 task 3. However, the scenario presented in this dataset is relatively simple for region-specific audio tagging, since the average number of overlapping sound events is 1.32. Thus, we use SpatialScaper \cite{roman2024spatial} to simulate a new dataset, named Spatial Region-Specific Audio Tagging (SRSAT). The room is randomly chosen from the given room set\footnote{https://github.com/iranroman/SpatialScaper}. The mean and standard deviation of the number of sound events are set to 25 and 3. Each audio clip is in the format of a tetrahedral microphone array, 60 seconds long, sampled at 24 kHz. We generate 1,000 audio clips for training, 60 audio clips for validation, and 60 audio clips for testing, respectively. There are 13 sound classes (female speech, male speech, clapping, telephone, laughter, domestic sounds, walk, door, music, musical instrument, water tap, bell and knock) from FSD50K \cite{fonseca2021fsd50k} in both the STARSS23 and SRSAT datasets. In each time step, the annotation contains the classes and positions (azimuth, elevation, and distance) of the sound events. 

\subsection{Implementation Details and Evaluation Metrics}
A 2-second audio clip is randomly extracted from the 60-second audio signals. We ensure that the extracted audio clip has at least one sound event. During training, we perform audio channel swapping data augmentation using the method proposed in \cite{wang2023four} to enlarge the dataset by a factor of eight. STFT is calculated with a window of 512 samples and a hop size of 256 samples. Both IPD and directional features are calculated with microphone pairs of $(0, 0), (0, 1), (0, 2)$ and $(0, 3)$ \cite{chen2018multi}. We use the CNN-14 variant of PANNs as the model backbone. Each model is trained using the Adam optimizer with a learning rate $10^{-5}$. The model is trained for 50 epochs with an early stop mechanism at a patience of 10 epochs. Apart from the features mentioned above, we explore the learning-based angular features shown in the left part of Figure \ref{distance}. A learnable embedding layer is used to extract features $E$ for different input angles which is jointly trained with PANNs. In addition, for spectral and spatial features, we explore spatial cue-augmented log-spectrogram features (SALSA) proposed in \cite{nguyen2022salsa}. We use the angular region of 60$\degree$ as the region of interest. For DF and learning-based features, we use the middle angle of the region to extract the features. For the FOV feature, we calculate each DF in a resolution of 5$\degree$. Following the previous work \cite{xu2017convolutional, gong2021ast}, we use both mean average precision (mAP) and equal error rate (EER) as the metric to evaluate the model performance. 

\section{Experimental Results and Discussions}
\label{exp_results}
\subsection{Impact of Features}
\begin{table}[tbp]
\caption{Experimental results for different features. Bold numbers indicate the best performance.}
\centering
\begin{tabular}{lllcc}
\toprule[1pt]
\textbf{Spectral} & \textbf{Spatial} & \textbf{Angle} & \textbf{mAP} ($\uparrow$) & \textbf{EER} ($\downarrow$) \\ \hline\specialrule{0em}{1pt}{1pt}
LPS               & IPD              & DF             & 0.473   & 0.253     \\ \hline\specialrule{0em}{1pt}{1pt}
LPS               & IPD              & FOV            & \textbf{0.485}   & 0.260     \\ \hline\specialrule{0em}{1pt}{1pt}
LPS               & IPD              & learned        & 0.416   & 0.260     \\ \hline\specialrule{0em}{1pt}{1pt}
LPS               & GCCPHAT          & DF             & 0.479   & \textbf{0.246}     \\ \hline\specialrule{0em}{1pt}{1pt}

SALSA             & SALSA            & DF             & 0.455    & 0.260    \\ \bottomrule[1pt]
\end{tabular}
\label{feature_impact}
\end{table}

We show the experimental results on SRSAT using different feature combinations for 60-degree angular region-specific audio tagging in Table \ref{feature_impact}. Experimental results show that using LPS and IPD with FOV features achieves the best performance of mAP. One possible reason is that FOV features can focus more on the desired area than other angle features, as DF and learning-based can only attend to a single azimuth. However, the FOV feature is more computationally expensive than DF. With a resolution of 5$\degree$, the FOV features require 72 times computational costs than DF. Thus, we used DF for the remaining experiments, given its competitive performance over the FOV feature. For other spatial features, GCCPHAT has competitive performance with IPD.


\subsection{Impact of Models}
The experimental results for different models are demonstrated in Table \ref{model_impact_angle} and it shows that PANNs achieve the best performance. AST and PSLA do not perform well giving lower mAP and higher EER than PANNs, which shows that PANNs can be adapted from traditional audio tagging to region-specific audio tagging effectively. The potential reason of AST not performing good is that the transformer-based method generally requires a large amount of data for good results \cite{dosovitskiy2020image}.

\begin{table}[tbp]
\caption{Experimental results for different models.}
\centering
\begin{tabular}{llcc}
\toprule
\textbf{Model} &\textbf{\# Params} & \textbf{mAP} ($\uparrow$) & \textbf{EER} ($\downarrow$) \\ \hline\specialrule{0em}{1pt}{1pt}
PSLA \cite{gong2021psla}     & 64.1M      & 0.384    & 0.283    \\ \hline\specialrule{0em}{1pt}{1pt}
AST \cite{gong2021ast}     & 22.8M      & 0.360  & 0.308      \\ \hline\specialrule{0em}{1pt}{1pt}
PANNs \cite{kong2020panns}    & 79.7M      & \textbf{0.473}  & \textbf{0.253}      \\ \bottomrule
\end{tabular}
\label{model_impact_angle}
\end{table}

\begin{figure}[tbp]
  \centering
  \includegraphics[width=\columnwidth]{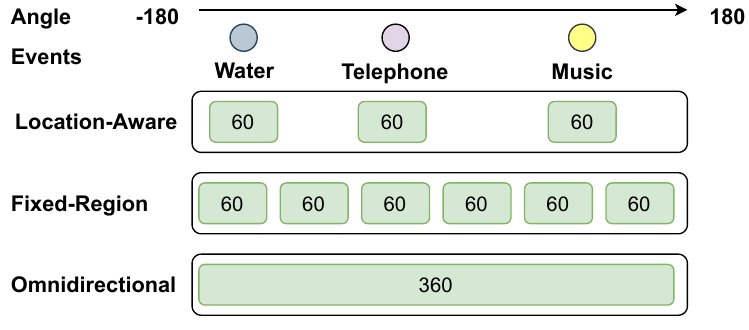}
  \caption{Illustration of different tagging system. The numbers are in degrees.}
  \label{tagging}
\end{figure}

\subsection{Compared with Omnidirectional Tagging Systems}

\begin{table}[tbp]
\caption{Experimental results for omnidirectional (OD), fixed-region (FR) and location-aware (LA) audio tagging systems.}
\centering
\scriptsize
\begin{tabular}{llllcc}
\toprule
\textbf{Spectral} & \textbf{Spatial} & \textbf{Angle} & \textbf{System} & \textbf{mAP} ($\uparrow$) & \textbf{EER} ($\downarrow$) \\ \hline\specialrule{0em}{1pt}{1pt}
LPS               & IPD              & -              & OD         & 0.371  & 0.267      \\ \hline\specialrule{0em}{1pt}{1pt}

LPS               & IPD              & FOV            & OD        & 0.370 & 0.274       \\ \hline\specialrule{0em}{1pt}{1pt}
LPS               & GCCPHAT          & -              & OD         & 0.376 & 0.263       \\ \hline\specialrule{0em}{1pt}{1pt}
SALSA             & SALSA            & -              & OD        & 0.373  & 0.265      \\ \hline\specialrule{0em}{1pt}{1pt}
LPS               & IPD              & DF             & FR    & 0.653 & 0.252         \\ \hline\specialrule{0em}{1pt}{1pt}
LPS               & IPD              & DF             & LA  & \textbf{0.667} & \textbf{0.244}         \\ \bottomrule
\end{tabular}
\label{general_tagging}
\end{table}

We compare the region-specific tagging systems with the systems for omnidirectional audio tagging, i.e., tagging sound events in the whole space instead of a specific region. We explore three systems as shown in Figure \ref{tagging}. The omnidirectional audio tagging system does not use the angle information. Both the location-aware system and the fixed-region system are based on a trained region-specific system. The location-aware system has the prior knowledge of the locations of all sound sources. The system is operated in the region of $60\degree$ centered at the azimuth of the sound events. We filter out overlapped regions, which are defined as two regions overlapping by $30\degree$, to reduce unnecessary computational cost. The fixed-region tagging system is operated on six fixed regions from $-180\degree$ to $180\degree$ at an interval of $60\degree$. 
For the location-aware and fixed-region systems, the system output and the ground truth from different regions are aggregated by the operation of maximum before calculating the metrics. 

The experimental results are shown in Table \ref{general_tagging}. Similar to the previous experiments, we compare the model performance achieved with different feature combinations for omnidirectional audio tagging. All omnidirectional systems have mAP between $0.37$ and $0.38$. It can be observed that adding the angle feature does not improve the model performance, as indicated by the third system. It is demonstrated that the fixed-region system outperforms the omnidirectional tagging one, which shows that the fixed-region system benefits from the region division. If the positions of the sound events are known, the model performance can be further improved with mAP from 0.653 to 0.667 and EER from 0.252 to 0.244.

\subsection{Impact of Angular Range}

\begin{table}[tbp]
\caption{Impact of angular range.}
\centering
\scriptsize
\begin{tabular}{lllcll}
\toprule
\textbf{Spectral} & \textbf{Spatial} & \textbf{Angle} & \textbf{Range} & \textbf{mAP} ($\uparrow$) & \textbf{EER} ($\downarrow$) \\ \hline\specialrule{0em}{1pt}{1pt}
LPS               & IPD              & DF             & 60\degree               & \textbf{0.473}   & \textbf{0.253}     \\ \hline\specialrule{0em}{1pt}{1pt}
LPS               & IPD              & DF             & 180\degree              & 0.406  & 0.266      \\ \hline\specialrule{0em}{1pt}{1pt}
LPS               & IPD              & DF             & 300\degree              & 0.370    & 0.275    \\ \bottomrule
\end{tabular}
\label{angular}
\end{table}

We explore the influence of different angular range and show the results in Table \ref{angular}. We can see that the proposed task becomes harder as the angular range becoming larger as more sound events would fall into the selected area, making the task more challenging. When the angular range is $300\degree$, mAP is 0.370, which is close to the model performance in the omnidirectional tagging scenario shown in Table \ref{general_tagging}. 

\subsection{Model Performance on STARSS23 Dataset}

\begin{table}[tbp]
\caption{Dataset statistics and performance comparisons between STARSS23 and SRSAT dataset.}
\centering
\begin{tabular}{lcc}
\toprule
                  & \multicolumn{1}{l}{\textbf{STARSS23}} & \multicolumn{1}{l}{\textbf{SRSAT}} \\ \hline\specialrule{0em}{1pt}{1pt}
Avg. No. Events &   1.320                                     & 2.233                             \\ \hline\specialrule{0em}{1pt}{1pt}
Max. No. Events   & 6                                         & 10                                \\ \hline\specialrule{0em}{1pt}{1pt}
Prop. One Event   & 0.623                                         & 0.287                                \\ \hline\specialrule{0em}{1pt}{1pt}
Prop. Two Events   & 0.278                                         & 0.294                                \\ \hline\specialrule{0em}{1pt}{1pt}
mAP of PANNs ($\uparrow$) & 0.938                                      & 0.473                            \\ \bottomrule
\end{tabular}
\label{dcase}
\end{table}
We further evaluate PANNs on the STARSS23 dataset and make comparisons with the model performance on SRSAT. The results are shown in Table \ref{dcase}. STARSS23 was originally used for sound source localization and detection. We can see that STARSS23 is much simpler than SRSAT for the proposed task, as indicated by the lower average number of events (Avg. No. Events) and maximum number of events (Max. No. Events) per frame. We show the proportions of frames containing one event (Prop. One Event) or two events (Prop. Two Events). It is demonstrated that STARSS23 dataset contains one event and two events at $90\%$ of the time frames, which necessitates the creation of SRSAT dataset. 

\subsection{Model Performance for a Given Distance}

\begin{table}[tbp]
\caption{Model performance for a given distance.}
\centering
\begin{tabular}{llcc}
\toprule
\textbf{Model} & \textbf{Param} & \multicolumn{1}{l}{\textbf{mAP} ($\uparrow$)} & \textbf{EER} ($\downarrow$) \\ \hline\specialrule{0em}{1pt}{1pt}
PANNs \cite{kong2020panns}         & 79.8M          & \textbf{0.417}      & \textbf{0.266}                      \\ \hline\specialrule{0em}{1pt}{1pt}
AST \cite{gong2021ast}           & 23.0M          & 0.324      & 0.320                      \\ \hline\specialrule{0em}{1pt}{1pt}
PSLA \cite{gong2021psla}          & 64.2M          & 0.399     & 0.274                       \\ \bottomrule
\end{tabular}
\label{distance_results}
\end{table}

In this part we explore the model performance for distance-guided audio tagging. We randomly select a sound event and use the ground truth distance as the condition. The performance of the model is reported in Table \ref{distance_results}. For each model, we use LPS as the spectral feature and IPD as the spatial feature following the previous settings. The distance feature extraction network is jointly trained with the model. The PANNs model achieves the best performance while AST and PSLA do not perform well. Compared with the model performance on azimuth-queried tagging reported in Table \ref{model_impact_angle}, distance-queried tagging scenarios are more challenging. 
One possible reason is that the statistic-based DF reflects the confidence of the sound source coming from a given azimuth, which can capture the characteristics of the selected regions better than the learning-based distance feature.

\section{Conclusion}

We have presented a novel task named region-specific audio tagging for spatial sound. We extended the current advanced tagging models for this task using angular region-queried and distance-queried mode. We further explored the impact of different combinations of spectral, spatial and positional features.
We find that using LPS and IPD combined with FOV features achieves the best mAP. 
When extending the region-specific tagging system to omnidirectional tagging, we find that the proposed fixed-regional and location-aware tagging system outperforms the omnidirectional tagging system. 
In the current setting, we only include 13 sound classes following the setting of DCASE Task 3. In future, we plan to include more sound classes from AudioSet to enhance the diversity of dataset.




\clearpage
\bibliographystyle{IEEEtran}
\bibliography{refs}







\end{document}